\def\sst{\scriptscriptstyle}
\def\sl{\slshape}

\def\kn#1{{\kern -#1 true cm}}

\newcommand{\ci}{\cite}

\def\bl{\renewcommand{\baselinestretch}} 
\def\cl{\centerline}
\def\bib#1{\bibitem[#1]{#1}}

\def\lp{\left(}
\def\rp{\right)}
\def\lb{\left[}
\def\rb{\right]}
\def\la{\langle\, }

\def\ra{\,\rangle} 

\def\LB{\left\lbrace}
\def\RB{\right\rbrace}

\def\+#1#2{{#1+#2}}

\def\top#1#2{\smash{\mathop{\hbox to .2cm{$#2$}}\limits^{#1}}}

\def\0#1{{(#1)}}
\def\1#1{{\hat #1}}
\def\2#1{{\tilde #1}}
\def\3#1{{\boldsymbol#1}}
\def\4#1{{\,\mathbb#1}}
\def\5#1{{\mathcal#1}}
\def\6#1{_{\sst#1}}
\def\7#1{{\bar#1}}
\def\8{\infty}
\def\9#1{^{\,\sst#1}}
\def\/#1{{\bf#1}}
\def\;#1{{\breve#1}}
\def\`#1{{\mathring#1}}

\def\^{\wedge}

\def\bh#1{{\boldsymbol{\hat{#1}}}}

\def\tc#1{\2{{\5#1}}}

\def\a{\alpha}

\def\d{\delta} 
\def\e{\varepsilon} 

\def\f{\phi}

\def\k{\kappa}

\def\o{\omega} 
\def\p{\pi} 

\def\q{\theta} 
 
\def\r{\rho}

\def\t{\tau}

\def\D{\Delta}

\def\Q{\Theta}

\def\hb#1{{\qq\text{#1}\qq}}
\def\bull{$\bullet\ $}
\def\db{{d\kern-.8ex {^-}}}

\def\diamond{$\diamondsuit$}
\def\heart{$\heartsuit$}
\def\spade{$\spadesuit$}
\def\club{$\clubsuit$}
\def\Star{$\bigstar$}

\def\={\equiv} 
\def\app{\approx} 
\def\cc#1{{{\mathbb C\hskip.5pt}^{#1}}}

\def\curl{\nabla\times} 
 
\def\eg{{\it e.g., }}

\def\ie{{\it i.e., }}

\def\im{{\,\rm Im}\ }  
\def\imp{\ \Rightarrow\ }
\def\i1#1{\int_{-\infty}^\infty d#1\, }

\def\lra{\leftrightarrow}

\def\lor#1{\4R^{#1,1}}
 
\def\mink{\4R^{3,1}}

\def\nn{{n+1}}
\def\pl{\partial}

\def\plra{\pl^{\kern-1.25ex^\lra}}
\def\qq{\quad}

\def\re{{\,\rm Re}\  }   
\def\rr#1{{{\mathbb R}^{#1}}}

\def\sr{\sqrt}

\def\supp{{\rm supp \,}}
\def\sv#1{\vskip#1ex}

\def\and{{\hbox{\  \ and\ \ }}}

\def\*{*\!}

\def\VE{\vfill\eject}

\def\bbx#1{ \boxed{\boxed{#1}}}

\def\bx#1{{ \boxed{\ #1\ }}}

\def\frame#1#2{
    \cl{\vbox{\hrule height .3pt
    \hbox{\vrule width .3pt\kern 5pt
    \vbox{\kern 5pt
    \vbox{\hsize #1cm\noindent#2}
    \kern 5pt}
    \kern 5pt\vrule width .3pt}
    \hrule height 0pt depth.3pt}}}

\documentclass[12pt]{article}  
\usepackage{amssymb,amsmath}  
\usepackage{graphicx}

\begin{document}     

{\tiny
\cl{Invited talk, Institute for Pure and Applied Mathematics:}
\cl{\sl Multiscale Geometric Analysis: Theory, Tools,  Applications\rm }
\cl{www.ipam.ucla.edu/programs/mga2003/}
}
\sv1

\cl{\large\bf Making Pulsed-Beam Wavelets}

\cl{\bf  Gerald Kaiser\footnote{Supported by AFOSR 
    Grant  \#F49620-01-1-0271.}}
	 \cl{The Virginia Center for Signals and Waves}
   \cl{kaiser@wavelets.com $\bullet$\  www.wavelets.com}
 
\sv2

\bl{1.5}

\bf \rm

\cl{\bf Abstract \rm}

Point sources in complex spacetime, which generate acoustic and electromagnetic pulsed-beam wavelets, are rigorously defined and computed with a view toward their realization.

\sv3

\bf 1. Introduction \rm

Acoustic and EM waves are natural candidates for \bf multiscale analysis\rm:  the 
wave/Maxwell equations have no \sl a priori \rm scale.  

This motivated the idea of  \bf physical wavelets \rm \ci{K94a, K02a},  localized solutions $W_z$  giving \bf frames \rm in solution spaces.

$z=x+iy$ is a  \bf complex spacetime point \rm generalizing the  time and scale parameters in 1D wavelets.  

$W_z$  \bf propagates \rm and can't have bounded spatial support. 

But its \bf source \rm does, giving $z$ a simple interpretation:  

\VE

\bull $x=(\3x,t)$  is the  \bf source  center \rm  in space and time

\bull $y=(\3y, u)$ gives its  \bf extension \rm around $x$: 

$\qq  \circ\  \3y$ gives  radius and orientation of  \bf launching disk \rm 

$\qq  \circ\   u$  gives  \bf pulse duration \rm 

\bull $(x,y)\sim $  4D version of \bf time \& scale\rm

\bull \bf Stability \rm  $\imp y$ is  timelike $\imp z\in\5T=$ \bf causal tube\rm.

Thus $W_z$ is a \bf  pulsed beam (PB) \rm with origin, direction, sharpness and duration controlled by $y$.

\bf Folklore: \rm In \bf\sl  some \rm sense,  \sv2

\cl{\frame{9.3}{$W_z\sim$  wave emitted by a \bf complex sourcepoint \rm $z$.}}

Since the 1970s,  \bf complex-source beams \rm have been used  in engineering; see \ci{HF01} for a comprehensive  review.  

Related ideas have circulated in relativity since the 1960s, \eg the \bf Kerr-Newman   \rm solution \ci{N65} representing charged,  spinning black holes; see also  \ci{K01a, N02}. 

My motivation came from a long-time project combining QM with relativity via coherent-state representations, with $\5T$ as \sl extended phase space \rm  \ci{K77, K78, K87, K90}. 

\bull  \bf PB wavelet  analysis of waves \rm - local alternative to Fourier.

\bull  Associated \bf complex sourcepoint analysis \rm of  sources.

\bull  Applications to radar and communications:  \ci{K96, K97, K01}. 
 
\bf \sl  Can platforms be made to launch and detect $W_z$'s?  \rm

Must understand the nature of the \bf sources\rm, making precise the notion of a wave emitted ``from''  $x+iy$. See also \ci{HLK00}.

\sv2

\bf 2. Point sources in complex space \rm

First define a point source in complex \bf space \rm  $\cc n, n\ge 3$.

($n=2$ is special and must be treated separately.)

The point source at $\3x=\30\in\rr n$ can be defined  as the source of the 
Newtonian potential $G_n$:
\begin{gather*}
\d_n(\3x)=\D_n G_n(\3x), \qq  G_n(\3x)=\frac1{\o_n}\, \frac{r^{2-n}}{2-n},
\end{gather*}
where $\o_n$ is the area of the unit sphere in $\rr n$ and 
\begin{align*}
r(\3x)=\sr{\3x\cdot\3x}=\sr{\3x^2}
\end{align*}
is the Euclidean distance.

Define the  \bf complex-distance \rm function 
\begin{align*}
\2r(\3x+i\3y)=\sr{(\3x+i\3y)^2}=\sr{r^2-a^2+2i\3x\cdot\3y},\qq a=|\3y|>0.
\end{align*}
Fixing $\3y\ne\30$, the branch points form an $(n-2)$-sphere
\begin{align*}
\5B(\3y)=\{\3x\in\rr n:  r=a, \   \3x\cdot\3y=0\}.
\end{align*}
The simplest branch cut for which $\2r(\3x)=+r(\3x)$ is
\begin{align*}
\re \2r\ge 0, \hbox{ which implies } -a\le\im\2r\le a.
\end{align*}
The \bf branch cut \rm is the $(n-1)$-disk spanning $\5B(\3y)$,
\begin{align*}
\5D(\3y)=\{\3x:  r\le a, \   \3x\cdot\3y=0\},\qq \pl\5D=\5B.
\end{align*}
We now  \bf define \rm the point source at $\3x=-i\3y$ by
\begin{align*}\bbx{ \ 
\2\d_n(\3x+i\3y)=\D_n  G_n(\3x+i\3y),\ \ 
G_n(\3z)\=\frac1{\o_n}\, \frac{\2r^{2-n}}{2-n} \ }
\end{align*}
where $\D_n$ is  the distributional Laplacian in $\3x$.

$\2\d_n(\3x+i\3y)$ is a \bf distribution \rm supported in $\3x\in\5D(\3y)$ \rm \ci{K00}. 

For \bf even \rm $n$, $G_n(\3z)$ is analytic whenever   $\3z^2\ne 0$, hence 
\begin{align*}
\supp_{\!\3x} \,\2\d_n(\3x+i\3y)=\5B(\3y)   \hbox{  for even } n \ge 4.
\end{align*} 
But for \bf odd \rm $n$, $G_n$ inherits the branch cut and
\begin{align*}
\supp_{\!\3x} \,\2\d_n(\3x+i\3y)=\5D(\3y)   \hbox{  for odd } n \ge 3.
\end{align*} 
The distribution $\2\d_3(\3z)$ will be computed along with the time-dependent one for pulsed beams in $\mink$.

\sv2

\bf 3.  Complex spacetime  sourcepoints \rm

Begin with  \bf Euclidean \rm $\rr4$  and complexify:
\begin{gather*}
x\6E=(\3x,-u),\   \   y\6E= (\3y, t ) \in\rr4\\
z=x\6E+iy\6E=(\3x+i\3y, -u+it) \in\cc4.
\end{gather*}
Rewrite this as a complex  \bf Minkowski \rm vector:
\begin{gather*}
z=x+iy=(\3z,i\t), \ \ \3z=\3x+i\3y,\  \t=t+iu\\
x=(\3x, it),\  y=(\3y, iu)\in\mink\\
x^2=\3x^2-t^2,\  \   y^2=\3y^2-u^2,\  \  z^2=\3z^2-\t^2.
\end{gather*}
Consider the Newtonian potential in $\rr 4$ and its extension,
\begin{align*}
 G_4 (x\6E)=-\frac1{4\p^2x\6E^2},\qq G_4\0z=-\frac1{4\p^2z^2}.
\end{align*}
$ G_4 (x\6E)$ is a fundamental solution for the Laplacian,
\begin{align*}
\D_4 G_4(\3x, u)=(\D_\3x+\pl_u^2)  G_4(\3x, u)=\d_4(\3x, u). \tag{$\5E$}
\end{align*}
\bf\sl Naive question\rm:  Does  $u\to it$ gives a fundamental solution for the \bf wave operator\rm,
\ie
\begin{gather*}
\square G_4(\3x, it)\=(\D_\3x-\pl_t^2) \,G_4(\3x, it) \sim \d (\3x,  t) \,??
\end{gather*}

\bf\sl Answer: \bf  No! \rm 

\bull $G_4$ is singular on $x^2=0$, so it must be \bf defined \rm  in $\mink$.

\bull When properly defined, it will be  \bf sourceless. \rm

\bull   \bf Causality \rm makes sense in $\4R^{3,1}$ but not in $\rr4$.

To find the correct definition, note that
\begin{align*}
-z^2&=\t^2-\3z^2=\t^2-\2r^2=(\t+\2r)(\t-\2r),
\end{align*}
which gives the partial-fractions decomposition
\begin{align*}\bbx{\ 
G_4\0z\!=G\9+\0z - G\9-\0z,\   
G\9\pm\0z= \frac {1}{8\p^2 \2r}\cdot \frac1{\t\mp\2r}\  } \tag{\spade}
\end{align*}
Since formal differentiation gives
\begin{align*}
\square G\9\pm\0z=0,
\end{align*}
$ G\9\pm\0z$ can have sources  only at $\3x\in\5D$  or  $\t=\pm\2r$.

Define the  \bf time-dependent radiation pattern \rm 
\begin{gather*} \tag{$\5R$}
2\p R\9\pm=\frac i{\t\mp\2r}=\frac i{(t\mp p)+i(u\mp q)}\\ 
\2r=p+iq ,\qq p>0,\  \bx{-a<q<a.}
\end{gather*}
Now $(p,q)$ are \bf  oblate spheroidal coordinates \rm in $\rr3$ whose level surfaces are $\5B$-\bf confocal  ellipsoids and hyperboloids\rm. 

In cylindrical coordinates with $x_3=\bh y\cdot\3x$ and $\r=\sr{r^2-x_3^2}$\,,
\begin{align*}
E_p&\=\{\hbox{constant } p>0\}=\LB\3x: \frac{\r^2}{a^2+p^2}+\frac{x_3^2}{p^2}=1\RB\\
H_q&\=\{\hbox{constant } q^2<a^2\}=\LB\3x: \frac{\r^2}{a^2-q^2}-\frac{x_3^2}{q^2}=1\RB.
\end{align*}
$E_p$'s are \bf wave fronts\rm, and $H_q$'s give orthogonal  \bf flow\rm. 

In the \bf far zone \rm $r\gg a$, they become \sl spheres and cones: \rm
\begin{gather*}
\2r=\sr{r^2-a^2+2iar\cos\q}\app r+ia\cos\q\\
 E_p\to\{r=p\},\qq H_q\to\{ \cos\q=q/a\}\\
\bbx{2\p R\9\pm\app \frac i{(t\mp r)+i(u\mp a\cos\q)}\ 
\top{t=\pm r}{\longrightarrow}\  \frac1{u\mp a\cos\q}.}\tag{\club}
\end{gather*}
\bf (I) $y$ is \bf timelike\rm:  $|u|>a$,  $G\9\pm$ is a smooth \bf  pulse \rm around $\pm\3y$  peaking at $t=\pm r$, with \bf duration \rm $T(\3x)=|u\mp a\cos\q|$ and  \bf elliptical \rm radiation pattern   of \bf  eccentricity \rm $a/|u|$.

\bf (II) $y$ is \bf spacelike\rm:  $|u|< a$, $G\9\pm$ is  singular  on the  cone  $\cos\q=\pm u/a$ at $t=\pm r$, with  a \bf hyperbolic \rm radiation pattern.

\bf (III) $y$ is \bf lightlike\rm:  $|u|=a$,  $G\9\pm$ is  singular  on \bf ray \rm  $\cos\q=\pm 1$ at $t=\pm r$, with  a \bf parabolic \rm  radiation pattern.

Only  \bf (I) \rm gives a reasonable PB,  though  \bf (II, III) \rm should be of interest otherwise since $G_4\0z$ is holomorphic for \sl all \rm $z^2\ne 0$.

We therefore \bf assume  \rm                 
\begin{align*}
y\in V\6\pm=\{(\3y, iu): \pm u> |\3y|  \}=\hbox{future/past cone},
\end{align*}
so that $z$ belongs to the \bf causal tube \rm $\5T=\5T\6+\cup\5T\6-$, where
\begin{align*}
\5T\6\pm =\{x+iy:  y\in V\6\pm\}=\hbox{future/past  tube}.
\end{align*}
$\5T\6\pm$  are famous in physics (quantum field theory, twistors) as well as mathematics
(Lie groups, harmonic analysis).

\bf Relation to propagators: \rm   Fix  $y\in V\6\pm$ and write
\begin{align*}
G^\k(x\pm i 0)&=\lim_{\e\to 0\9+}G^\k(x\pm i\e y), \qq  \k=+,\  -.
\end{align*}
Then (\spade) gives (with $\5P$= principal value)
\begin{align*}
G\9\pm(x+i0)&=\frac 1{8\p^2 r}\,\5P\frac1{t\mp r}-\frac {i\d(t\mp r)}{8\p r}\\
G\9\pm(x-i0)&=\frac 1{8\p^2 r}\,\5P\frac1{t\mp r}+\frac  {i\d(t\mp r)}{8\p r}.
\end{align*}
\bf Huygens' principle \rm  requires the \bf Minkowskian limit \rm
\begin{gather*} \tag{$\5M$}
 G\6M\9\pm\0x=G\9\pm(x-i0)-G\9\pm(x+i0)=\frac {i\d(t\mp r)}{4\p r}.
\end{gather*}
These are the \bf retarded and advanced  propagators\rm,
fundamental solutions vanishng for $\pm t<0$:
\begin{align*}
\square  G\6M\9\pm\0x=- i\d(\3x)\d\0t=\d\0x, \tag{$\5H$}
\end{align*}
where the last equality comes from our volume element,
\begin{align*}
x=(\3x, it)\in \4R^{3,1}\imp dx=id\3x dt.
\end{align*}
\bull ($\5H$) is the desired hyperbolic counterpart of ($\5E$). 

\bull $G\6M=G\6M\9+-G\6M\9-$ is the \bf Riemann function\rm:
\begin{gather*}
G\6M\0x=\frac {i\d(t-r)}{4\p r}-\frac {i\d(t+ r)}{4\p r}\\
\square G\6M\0x=0,\ \  G\6M(\3x,0)=0,\ \  \pl_{it}G\6M(\3x,0)=\d(\3x).
\end{gather*} 
\bull \bf Causality \rm  comes with a \bf choice of branch \rm and has no meaning for $G_4\0z$:  $\2r\to-\2r\!\!\imp\!\! G\9\pm\to -G\9\mp$.
\sv1

$(\5M)$ is typical of \bf hyperfunction theory\rm, representing
distributions as limits of differences of \sl local \rm  holomorphic functions  (in general, sheaf cohomology classes)  \ci{K88}. 

We now define  the \bf  point source at $x=-iy$ \rm as
\begin{align*}\boxed{\ 
\2\d\0z=\square \, G\9\pm\0z,\  \ z=x+iy.   \  }\tag{\Star}
\end{align*}

\bull This is identical for  $G\9+$  and $G\9-$:
\begin{align*}
 z^2=0\imp x^2=y^2<0 \hbox{ and } x\cdot y=0,
\end{align*}
but $x\cdot y\ne 0$ since both vectors are timelike

Hence  $G_4$ is holomorphic in $\5T$ and
\begin{align*}
\square\, G\9+\0z -\square\, G\9-\0z=\square\, G_4\0z=0.
\end{align*}
\bull By ($\5H$),  the Minkowskian limit of $\2\d$ is $\d$:
\begin{align*}
\d\6M\0x\=\2\d(x-i0)-\2\d(x+i0)=\d\0x. \tag{$\5M'$}
\end{align*}
\bull   Since $\t\mp\2r\ne 0$ in $\5T$, $G\9\pm$ is singular only when $\3x\in\5D$
and
\begin{align*}
\supp \2\d(x+iy)=\{(\3x, it): \3x\in\5D(\3y)\}\=\tc D(\3y) \qq \forall y\in V\6\pm\,.
\end{align*}
$\tc D(\3y)$ is the  \bf world tube \rm swept out in $\mink$ by $\5D(\3y)$ at rest.

\VE

\bf 4. Computing the source distributions \rm

Consider the function  on $\5T$ defined by
\begin{align*}
W(\3z, i\t)=\frac{g(\t-\k\2r)}{4\p\2r},\qq \k= \pm \tag{$\5W$}
\end{align*}
with   $g(\t)$  an \bf analytic signal\rm, holomorphic for $\im\t\ne 0$.

We want to compute the source distribution
\begin{align*}
S\0z=\square_x W\0z.
\end{align*}
Note that 
\begin{align*}
g(\t)=\frac 1 {2\p\t}&\imp W\0z= G^\k\0z\imp S\0z=\2\d\0z\\
g(\t)\=-1&\imp W\0z= G_3(\3z)\imp S\0z=\2\d(\3z).
\end{align*}
\bull  $\im(\t\pm \2r)\ne 0$ in $\5T\imp\supp S\subset \tc D(\3y)$. 

\bull $W$ is \bf doubly singular \rm on $\tc D$, where $g(\t-\k\2r)$ has a  jump.

Regularize $W$ with the \sl Heaviside function \rm  $\Q$. 
Let $\e>0$ and
\begin{align*}
W_\e\0z=\Q(p-\e) W\0z=\begin{cases}
    W\0z  & \text{outside } E_\e\\
    0  & \text{inside } E_\e\,.
\end{cases}
\end{align*}
Then the \sl regularized  source, \rm defined by
\begin{align*}
S_\e\0z\=\square W_\e\0z,
\end{align*} 
is supported on the world tube  $\2E_\e\subset\mink$ of $E_\e$. 

A computation gives
\begin{align*}
 |\2r|^2 S_\e=\d' \frac{(p^2+a^2) g}{4\p  \2r}
+\d\frac{(\e^2+a^2)(\2r g_p-g)}{2\p  \2r^2}+\d\frac{\e g}{2\p\2r}\,,
\end{align*}
where
\begin{align*}
\d(p-\e)&=\Q'(p-\e), \qq g_p\=\pl_p g=-\k g'(\t-\k\2r).
\end{align*}
$S_\e$ acts on a test function $f(\3x)$ (no smearing needed in $t$) by
\begin{gather*}
\la S_\e\,, f\ra\= \int d\3x\ S_\e(\3x+i\3y, i\t) f(\3x)\\ 
\hb{with} d\3x=\frac{p^2+q^2}a\, dp\,dq\, d\f. \tag{Vol}
\end{gather*}

Integrating by parts in $p$ and simplifying gives
\begin{align*}
\la S_\e\,, f\ra&=\frac{\e^2+a^2}{2a}\int_{-a}^a dq \lb \frac{g_p\`f}{\2r}
-\frac{g\`f}{\2r^2}-\frac{g\`f_p}{\2r} \rb_{p=\e}
\end{align*}
where $\`f(p,q)$ is the \sl mean \rm of $f(p,q,\f)$ over $\f$.

But $g_p=-i g_q$ as $g$ is holomorphic. Integrating by parts in $q$,
\begin{gather*}
\la S_\e\,, f\ra=\frac{\a\7\a}{2ia}\lb \frac{g\`f}{\2r}\rb_{\2r=\7\a}^{\2r=\a}
\!\!-\frac{\a\7\a}{a}\!\int_{-a}^a \!\!\!dq\ \frac{g \`f_\2r}{\2r}\!\Bigm|_{p=\e} \tag{\diamond}
\end{gather*}
with $\a=\e+ ia$ and $\7\a$ the \bf north and south poles \rm of $E_\e$,  $\`f(\2r)\=\`f(p,q)$ (no analyticity implied in $p+iq$) and
\begin{align*}
 \`f_\2r(\2r)=\pl_\2r \`f(\2r)\=\frac12 (\pl_p-i\pl_q)\`f(p,q).\tag{$\`f_\2r$}
\end{align*}
Taking the limit $\e\to 0$ now gives the action of $S$:
\begin{align*}\tag{\heart}
\la S, f\ra=-\1g(\t,a) f(\30)+2ia\!\!\int_0^a \!\frac{dq}q\ \1g(\t,q)\`f_\2r(iq)\Bigm|_{p=\e}  
\end{align*}
where
\begin{align*}
\1g(\t, q)=\frac12\lb g(\t-i q)+g(\t+i q) \rb
\end{align*}
is the average of $g(\t-\k\2r)$ over the jump at $\2r=\pm iq\in\5D$ and we
used the continuity of $f$ and its derivatives across $\5D$,
\begin{gather*}
\`f(-iq)=\`f(iq),\qq \`f_\2r(-iq)=-\`f_\2r(iq)\\
 \`f(\pm ia)= f(\30),\qq \`f_\2r\00=0,
\end{gather*}
which ensures that the integral in (\heart) is defined.

As a check, note that (\heart) gives  the correct value as $\3y\to\30$:
\begin{align*}
 \la S, f\ra\to-g(\t)f(\30)\imp S(\3x,i \t)=-g(\t)\d(\3x).
\end{align*}

\VE

\bf 5. Interpretation \rm

Let us examine the meaning of the right side in
\begin{gather*}
\la S_\e\,, f\ra=\frac{\a\7\a}{2ia}\lb \frac{g\`f}{\2r}\rb_{\2r=\7\a}^{\2r=\a}
\!\!-\frac{\a\7\a}{a}\!\int_{-a}^a \!\!\!dq\ \frac{g \`f_\2r}{\2r}\!\Bigm|_{p=\e}. \tag{\diamond}
\end{gather*}
\bull The boundary term is a pair of  \bf real  sourcepoints at the poles \rm 
$\2r=\a, \7\a$ ($\r=0,\  x_3=\pm\e$) of $E_\e$  modulated by $g$.

\bull Since $\pl_p$ is outward-orthogonal to $E_\e$,  the term with $\pl_p\`f$ is a \bf double layer \rm  on $E_\e$, modulated by $g$; see Eq. ($\`f_\2r$).

\bull $\pl_q$ is tangent to $E_\e$, representing a  flow from $\7\a$ to $\a$. Hence   the term with $\pl_q\`f$ is a \bf flow \rm on $E_\e$  modulated by $g$.

\bull Since $g\0\t$ is an  analytic signal \ci{K94}, the modulation by
\begin{align*}
g=g(\t\mp \2r)=g(t\mp \e+i(u\mp q))
\end{align*}
is  strongly \sl amplified if $\pm qu>0$ and  diminished \rm  if
$\pm qu<0$. This makes $W\0z$ a  \bf unidirectional beam\rm.

Taking into account (Vol)  gives the \sl unsmeared \rm form
\begin{gather*}\bbx{\ 
S_\e\0z=W\0z \!\lb i\d(\2r-\7\a) - i\d(\2r-\a) 
-2\Bigm| \frac{\a}{\2r}\Bigm|^2\!\!\d(p-\e)\,\pl_\2r \rb  \ }
\end{gather*}
where
\begin{align*}
\d(\2r-\e\pm ia)=\d(p-\e)\d(q\pm a).
\end{align*}

Similarly, (\heart) gives the unsmeared form 
\begin{align*} 
S\0z=-\1g(\t,a) \d(\3x)-\1g(\t, q)\frac{a^2\d\0p}{2\p i q^3}\,\pl_\2r\,.
\end{align*}
In cylindrical coordinates, this becomes
\begin{align*}\bbx{\ 
S\0z=-\2g(\t, 0)\d(\3x)-\2g(\t,\r)\,\frac{\Q(a-\r)\, \d(x_3)}{2\p\sr{a^2-\r^2}}\,
\Bigl( \frac a\r\, \pl_\r-i\pl_3\Bigr)\ }
\end{align*} 
where  
\begin{align*}
\2g(\t,\r)=\frac12\lb g(\t-i\sr{a^2-\r^2}) +g(\t+i\sr{a^2-\r^2})\rb.\end{align*}

Letting $g\=-1$  gives the static complex sourcepoint
\begin{align*}\bbx{\ 
\2\d(\3z)=\d(\3x)+ \frac{\Q(a-\r) \d(x_3)}{2\p\sr{a^2-\r^2}}
 \lp\frac a\r\,\pl_\r-i\pl_3\!\rp \ }
\end{align*}
which is simpler than the form derived in \ci{K00}, being  \sl local \rm while the latter had a subtraction.

Letting $g\0\t=1/2\p\t$ gives the complex spacetime point source,
\begin{align*}\bbx{\ 
\2\d\0z=\frac{\t \d(\3x)}{2\p z^2}+\frac{\t\Q(a-\r) \d(x_3)}{2\p z^2\sr{a^2-\r^2} } \,
\Bigl( \frac a\r\, \pl_\r-i\pl_3\Bigr),\ }
\end{align*}
where 
\begin{align*}
 z=(\3z, i\t)\in\tc D\imp z^2=\r^2-a^2-\t^2\ne 0.
\end{align*}

\bull The above method works in  $\lor n$ for all  \sl odd \rm $n\ge 3$.

\bull An interesting connection has been obtained between the \sl Euclidean \rm source distribution $\2\d_\nn$  and solutions of the \sl homogeneous \rm wave equation in $\lor n$, based on the fact that the latter are \bf spherical means \rm over $(n\!-\!1)$-spheres and $\2\d_\nn$ is supported precisely on these spheres for odd  $n$ \ci{K00}.

\bull These results extend to  \sl even \rm $n\ge 2$ by a variant of  Hadamard's \sl method of descent \rm  relating  $\2\d_n$  to $\2\d_\nn$ and $\2\d_{n,1}$ to $\2\d_{\nn,1}$. 

The cases $\2\d_2$ and $\2\d_{1,1}$ are special since $G_2$ is logarithmic, but the results are similar.

\sv2

\bf 6. Making waves \rm

 \bf Acoustic  \rm wavelets   are generated from a   `mother'   PB $W\9\pm$  by
\begin{align*}
W\9\pm_z(x')=W\9\pm(x'-\7z)=W\9\pm(x'-x+iy). 
\end{align*}

For suitable $g$,  subfamilies of  $W\9\pm_z$'s are \bf frames\rm,  giving a   \sl pulsed-beam analysis-synthesis  \rm scheme  for general waves as a \sl local \rm alternative to Fourier analysis. 

In \ci{K94}, the $W\0z$'s are (fractional) time derivatives of $G\9\pm\0z$. 

\VE

\bf Electromagnetic wavelets \rm   are frames of \bf\sl dyadic \rm PB propagators   $\4G\9\pm_z(x')=\4G\9\pm(x'-\7z)$ constructed as follows \ci{K94, K02}.

\/1. An \sl electric \& magnetic (E\&M)  dipole \rm with   \sl dipole moments \rm $\3p_e, \3p_m$ at  $x=-iy$ gives an  E\&M \sl 
\sl polarization density \rm
\begin{align*}
\3P\0z\=\3P_e\0z+i\3P_m\0z=\3p\,\2\d\0z,\qq \3p=\3p_e+i\3p_m\in\cc3.
\end{align*}
\/2. This gives retarded/advanced  E\&M \bf PB Hertz potentials \rm
\begin{gather*}
\3Z\9\pm\0z\=\3Z_e\9\pm\0z+i\3Z_m\9\pm\0z=\3p\, G\9\pm\0z\\
\square\3Z\9\pm\0z=\3p\,\square G\9\pm\0z=\3P\9\pm\0z.
\end{gather*}
\/3. These in turn generate  EM fields $\3F\9\pm\!\!=\3E\9\pm\!\!+i\3B\9\pm\!$ given by
\begin{align*}
\3F\9\pm\0z=\curl\curl\3Z\9\pm\0z-\pl_{it}\curl\3Z\9\pm\0z.
\end{align*}
\/4.  The dyadic `mother'  PBs  $\4G\9\pm$ are now defined by
\begin{align*}
\4G\9\pm\0z\3p=\3F\9\pm\0z.
\end{align*}
\bull \sl All the fields are  real\rm!  The \bf  self-dual  pairings \rm $\3E+i\3B$, etc. merely exhibit the  (complex!)  structure of EM. 

\bull Gives local analysis/synthesis schemes for fields and sources.

\bull PB analysis seems more natural for EM than scalar waves since the real and imaginary parts have  direct interpretations.

\sv1

Applications to radar and communications have been proposed \ci{K96, K97, K01} whose utility would be greatly enhanced if sources can be constructed to launch and detect EM wavelets.

\sv4

\small
\bl {1}

\bf Acknowledgements \rm

I thank Iwo Bialynicki-Birula, Ted Newman, Ivor Robinson and Andrzej Trautman for helpful discussions, and  Dr.~Arje Nachman for his sustained support of my work through the Air Force Office of Scientific Research under Grant  
\#F49620-01-1-0271.

\sv2

\end{document}